\documentclass[twocolumn,prb]{revtex4}
\usepackage{bm}
\usepackage{graphicx}
\newcommand{\nix}[1]{}

\begin{document}
\title{
Weak antilocalization in high-mobility two-dimensional systems}
\author{L.\,E.~Golub}
\affiliation{A.\,F.~Ioffe Physico-Technical Institute, Russian
Academy of Sciences, 194021 St.~Petersburg, Russia}


\begin{abstract}
Theory of weak antilocalization is developed for high-mobility
two-dimensional systems. Spin-orbit interaction of Rashba and
Dresselhaus types is taken into account. Anomalous
magnetoresistance is calculated in the whole range of classically
weak magnetic fields and for arbitrary strength of spin-orbit
splitting. Obtained expressions are valid for both ballistic and
diffusive regimes of weak localization. Proposed theory includes
both backscattering and nonbackscattering contributions to the
conductivity. It is shown that magnetic field dependence of
conductivity in high-mobility structures is not described by
earlier theories.
\end{abstract}


\maketitle

\section{Introduction}
Anomalous magnetoresistance caused by weak localization is a
powerful tool for extracting kinetic and band-structure parameters
of three-dimensional (3D) and 2D systems.\cite{AA} Theoretical
expression for magnetoconductivity valid in the whole range of
classically weak magnetic fields taking into account all
interference processes has been first derived in
Ref.\onlinecite{Zyuzin}. In the absence of spin-orbit interaction
the sign of the magnetoresistance is negative.

However the anomalous magnetoresistance is an alternating function
in 2D semiconductor systems. In particular, in low fields it is
positive and cannot be described by the theory
Ref.\onlinecite{Zyuzin}. The reason for positive magnetoresistance
is a spin-orbit interaction. In semiconductor heterostructures it
is described by the following Hamiltonian
\begin{equation}\label{H_SO}
    H(\bm{k}) = \hbar  \: \bm{\sigma} \cdot
    \bm{\Omega}(\bm{k}),
\end{equation}
where $\bm{k}$ is the electron wave vector, $\bm{\sigma}$ is the
vector of Pauli matrices, and $\bm{\Omega}$ is an odd function of
$\bm k$. The spin splitting due to spin-orbit interaction
Eq.~(\ref{H_SO}) equals to $2 \hbar \Omega({\bm k})$.

In the presence of spin splitting, weak magnetic field decreases
conductivity. Therefore the effect in such systems is called
\emph{weak antilocalization}. Theory of magnetoresistance in
systems with spin-orbit interaction Eq.(\ref{H_SO}) was developed
in Ref.\onlinecite{ILP}. However the obtained expressions are
valid only for i)weak spin-orbit interaction and ii)very low
magnetic fields. The first assumption means that $\Omega \tau \ll
1$, where $\tau$ is the scattering time. The second condition
reads as $l_B \gg l$, where $l_B = \sqrt{\hbar / e B}$ is the
magnetic length, and $l$ is the mean free path. This so-called
``diffusion'' regime takes place in fields $B \ll B_{tr}$, where
$$B_{tr} = {\hbar \over 2 e l^2}$$ is the ``transport'' field.

In high-mobility structures both these conditions fail. Due to
long scattering times the product $\Omega \tau$ can be even larger
than unity.\cite{Harley,WAL&SIA,WAL_Yulik_PRL,Hamburg} Besides,
the transport field is often less than
1~mT,~\cite{WAL&SIA,WAL_Yulik_PRL,Stud} that is too small range of
magnetic fields. This means that particle motion is rather
ballistic than diffusive. Therefore fitting experimental data by
the theories Refs.\onlinecite{Zyuzin,ILP} is not always
successful.\cite{Stud}

An attempt to derive the field dependence of anomalous
magnetoresistance for high-mobility structures has been performed
in Ref.\onlinecite{WAL_Yulik_PRL}. However the developed theory is
correct only for high fields $B/B_{tr} \gg (\Omega \tau)^2$ and
ignores some contributions to the conductivity.

The aim of the present work is to develop the
weak-antilocalization theory for systems with strong spin-orbit
interaction valid for both ballistic and diffusion regimes. The
magnetic field dependence of the conductivity is calculated for
arbitrary values of $B/B_{tr}$ and $\Omega \tau$, opening a
possibility to describe anomalous magnetoresistance experiments
and to extract spin-splitting and kinetic parameters of
high-mobility 2D systems.

\section{Theory}
There are two $\bm k$-linear contributions to the spin-orbit
interaction Hamiltonian Eq.~(\ref{H_SO}) in 2D semiconductor
systems: the Rashba term $\bm{\Omega}_R$ and the Dresselhaus term
$\bm{\Omega}_D$. In heterostructures grown along the direction $z
\parallel [001]$ both vectors $\bm{\Omega}_{R,D}$ lie in the 2D plane and have
the following form
\begin{equation}\label{Omegas}
\bm{\Omega}_R = \Omega_R \, (\sin{\chi}, -\cos{\chi}), \:\:\:
\bm{\Omega}_D = \Omega_D \, (\cos{\chi}, -\sin{\chi}).
\end{equation}
Here the axes are chosen as $x \parallel [100]$, $y \parallel
[010]$, and $\tan{\chi} = k_y/k_x$. The anomalous
magnetoresistance is the same if one takes into account the Rashba
or the Dresselhaus contribution. Therefore we consider below only
one term in $\bm \Omega$ with an isotropic spin splitting $2
\hbar\Omega \sim k$.

Retarded and advanced Green functions of a system with the
spin-orbit interaction Eqs.~(\ref{H_SO}),~(\ref{Omegas}) are $2
\times 2$ matrices in the spin space. In the Landau gauge under
scattering from a short-range potential, they are given
by\cite{STNS}
\begin{equation}\label{G}
    G^{R,A}(\bm{r},\bm{r}') = \sum_{N q s}
{\Psi_{Nqs} (\bm{r}) \Psi_{Nqs}^\dag (\bm{r}') \over E_{\rm F} -
E_{Ns} \pm {\rm i} \hbar /2 \tau \pm {\rm i} \hbar /2 \tau_\phi}.
\end{equation}
Here $E_{\rm F}$ is the Fermi energy, $N$ is a number of the
Landau level, $q$ is the wave vector in the 2D plane, $\tau_\phi$
is a phase relaxation time, $s = \pm$ enumerates two spin states,
$E_{Ns}$ is the electron energy, and two-component spinors
$\Psi_{Nqs}$ are the electron wave functions in the presence of
magnetic field and the spin-orbit interaction Eq.~(\ref{H_SO}).
For Rashba spin-splitting, $\Psi_{Nqs}$ is a superposition of the
electron states $| N, q, \uparrow \rangle$ and $| N+1, q,
\downarrow \rangle$,\cite{Rashba} i.e. with the same $N+s_z$,
where $s_z$ is the spin projection onto the growth axis. The
Dresselhaus spin-orbit interaction mixes the states with equal
$N-s_z$.

In low magnetic fields
\begin{equation}\label{field_range}
    \omega_c \ll \Omega, \: \tau^{-1} \ll E_{\rm F}/\hbar,
\end{equation}
where $\omega_c$ is the cyclotron frequency, one can show that
both magnetic field and spin-orbit interaction result in an
appearance of phases in the Green functions
\begin{equation}\label{G_phases}
    G^{R,A}(\bm{r},\bm{r}') = G^{R,A}_0(R)
    \: \exp{\left[ {\rm i} \varphi(\bm{r},\bm{r}') - {\rm i} \tau  \bm{\sigma} \cdot
    \bm{\omega}(\bm{R}) \right]}.
\end{equation}
Here $\bm{R} \equiv (X,Y) = \bm{r} - \bm{r}'$, $G^{R,A}_0(R)$ are
the Green functions at $B=0$ and $\Omega = 0$, and
$\varphi(\bm{r},\bm{r}') = (x+x')(y'-y)/2l_B^2$. The vectors
$\bm{\omega}(\bm{R})$ are determined by the symmetry of a
spin-orbit interaction\cite{LK}
\begin{equation}\label{omega}
    \bm{\omega}_R(\bm{R}) = {\Omega_R \over l} (Y, -X),
    \hspace{0.7cm} \bm{\omega}_D(\bm{R}) = {\Omega_D \over l} (X,
    -Y).
\end{equation}

The weak-localization correction to conductivity is determined by
interference of paths passing by a scattering particle in opposite
directions. The amplitude of this interference, Cooperon, depends
on four spin indices: ${\cal
C}_{\alpha\beta,\gamma\delta}(\bm{r},\bm{r}')$. Here $\alpha$ and
$\beta$ ($\gamma$ and $\delta$) are the spin states of a particle
before and after passing the path between the points $\bm{r}$ and
$\bm{r}'$ in the 2D plane forward (backward). The Cooperon
satisfies the matrix equation
\begin{equation}\label{C_Eq}
{\cal C}(\bm{r},\bm{r}') = {\hbar^3 \over m \tau}
P(\bm{r},\bm{r}') +
    \int d \bm{r}_1 P(\bm{r},\bm{r}_1) {\cal C}(\bm{r}_1,\bm{r}'),
\end{equation}
where $m$ is the electron effective mass and
$$P_{\alpha\gamma,\beta\delta}(\bm{r},\bm{r}') =
{\hbar^3 \over m \tau} G^{R}_{\alpha\beta}(\bm{r},\bm{r}')
G^{A}_{\gamma\delta}(\bm{r},\bm{r}')$$ is the probability for an
electron to propagate from $\bm{r}$ to $\bm{r}'$ forward and
backward.\cite{WAL_Yulik_PRL} It follows from Eq.~(\ref{G_phases})
that
\begin{equation}\label{P_phases}
P(\bm{r},\bm{r}') = P_0(R)\: \exp{\left[ 2 {\rm i}
\varphi(\bm{r},\bm{r}') - 2{\rm i} \tau \bm{S} \cdot
\bm{\omega}(\bm{R}) \right]},
\end{equation}
where $\bm{S}_{\alpha\gamma,\beta\delta} =
(\bm{\sigma}_{\alpha\beta} + \bm{\sigma}_{\gamma\delta})/2$ is an
operator of the total angular momentum of two interfering
particles, and $$P_0(R) = {\exp{(-R/\tilde{l})} \over 2 \pi R l}$$
is the value of $P$ in the absence of a magnetic field and a
spin-orbit interaction. Here the effective scattering length
$\tilde{l} = l/(1 + \tau / \tau_\phi)$.

In order to find the Cooperon, we expand the matrix $P$ into the
series over wave functions of a spinless particle with the charge
$2e$ in a magnetic field
\begin{equation}\label{P_expand}
    P(\bm{r},\bm{r}') = \sum_{N N' q} P(N,N') \: \Phi_{Nq}(\bm{r})
    \Phi_{N'q}^*(\bm{r}').
\end{equation}
The expansion coefficients are given by
$$P(N,N') = \int d \bm{r} \: P_0(r)\exp{[ - 2{\rm i} \tau \bm{S} \cdot
\bm{\omega}(\bm{r})]} F_{NN'}(\bm{r}),$$
where
$$F_{NN'}(\bm{r}) = e^{-t^2 / 2} \: L_N^{N'-N}(t^2) \: (-t e^{{\rm i} \phi})^{N'-N} \: \sqrt{N! \over N'!}.$$
Here $t=r/l_B$, $\tan \phi = y/x$, and $L_N^M$ are the associated
Laguerre polynomials. At $\Omega = 0$, $P(N,N') \sim
\delta_{NN'}$. Finite spin splitting leads to nonzero values of
$P(N,N')$ with $|N-N'| \leq 2$.

Expanding the Cooperon ${m \tau \over \hbar^3} {\cal C} (\bm{r},
\bm{r}')$ in series~(\ref{P_expand}) as well, we obtain the
following infinite system of linear equations for its expansion
coefficients
\begin{equation}\label{C_coeff}
    {\cal C}(N,N') = P(N,N') + \sum_{N_1} P(N,N_1) {\cal
    C}(N_1,N').
\end{equation}
In order to solve this system, we turn to the representation of
total angular momentum of two particles $\bm S$: $\alpha \gamma
\to S m_s$, where $S=0,1$ is the absolute value of $\bm S$, and
$m_s$ is its projection onto the $z$ axis ($|m_s| \leq S$). The
pair of particles with $S=0$ is in the singlet state while $S=1$
corresponds to the triplet one.

Spin-orbit interaction Eq.~(\ref{H_SO}) with only one contribution
Eq.~(\ref{Omegas}) remains the energy spectrum isotropic.
Therefore the particles being in the singlet and triplet states do
not interfere, and there are two uncoupled Cooperons corresponding
to the triplet and singlet, ${\cal C}_T$ and ${\cal C}_S$.

The singlet part is independent of the spin splitting and can be
found from Eq.~(\ref{C_coeff}) as for $\Omega = 0$:
\begin{equation}\label{C_S}
    {\cal C}_S(N,N') = {P_N \over 1-P_N} \delta_{NN'},
\end{equation}
where
$$    P_N = {l_B \over l} \int\limits_0^\infty dx
    \exp{\left( -x {l_B \over \tilde{l}} - {x^2 \over 2}\right)}
    L_N(x^2).
$$

The triplet part ${\cal C}_T(N,N')$ satisfies Eq.~(\ref{C_coeff})
with an infinite matrix $P_T(N,N')$. ${\cal C}_T$ and $P_T$ are
matrices with respect to both $N,N'$ and $m_s, m_s' =1,0,-1$. It
is crucial that $P_T$ can be decomposed into $3 \times 3$ blocks.
For Rashba spin-orbit interaction, it takes place in the basis of
the states $|N, m_s\rangle$ with equal $N+m_s$: $|N-2, 1\rangle$,
$|N-1, 0\rangle$, $|N, -1\rangle$, while for Dresselhaus term this
takes place for the states with the same $N-m_s$. In both cases
the blocks in $P_T$ can be obtained by a unitary transformation
from the following matrix
\begin{equation}\label{A_N}
    A_N = \left(%
\begin{array}{ccc}
  P_{N-2} - S_{N-2}^{(0)} & R_{N-2}^{(1)} & S_{N-2}^{(2)} \\ \nonumber
  R_{N-2}^{(1)} & P_{N-1} - 2S_{N-1}^{(0)} & R_{N-1}^{(1)} \\ \nonumber
  S_{N-2}^{(2)} & R_{N-1}^{(1)} & P_N - S_N^{(0)} \\
\end{array}%
\right).
\end{equation}
Here
\begin{eqnarray}\label{S_N}
    &&S_N^{(m)}= {l_B \over l} \sqrt{N! \over (N+m)!}    \nonumber\\
    &\times& \int\limits_0^\infty dx
    \exp{\left( -x {l_B \over l} - {x^2 \over 2}\right)}
    x^m L_N^m(x^2) \sin^2\left(\Omega \tau {l_B \over l} x \right),
    \nonumber
\end{eqnarray}
\begin{eqnarray}\label{R_N}
    &&R_N^{(m)}= {l_B \over l \sqrt{2}} \sqrt{N! \over (N+m)!} \nonumber\\
    &\times& \int\limits_0^\infty dx
    \exp{\left( -x {l_B \over l} - {x^2 \over 2}\right)}
    x^m L_N^m(x^2) \sin\left(2\Omega \tau {l_B \over l} x \right).
    \nonumber
\end{eqnarray}

The triplet part of the Cooperon is expressed via the
matrix~(\ref{A_N}) as follows: it consists of the blocks ${\cal
C}_T(N)$ given by
\begin{equation}\label{C_T}
    {\cal C}_T(N) = A_N (I-A_N)^{-1},
\end{equation}
where $I$ is a $3 \times 3$ unit matrix.

The conductivity correction due to weak antilocalization is given
by a sum of two terms\cite{Zyuzin}
$$\sigma(B) = \sigma_a + \sigma_b,$$
where $\sigma_a$ and $\sigma_b$ can be interpreted as
backscattering and nonbackscattering interference corrections to
conductivity.\cite{DKG} They are given by
\begin{eqnarray}\label{sigma_a_def}
    \sigma_a = {\hbar \over 4 \pi} \int d {\bm r} \int d {\bm r}'
    \sum_{\alpha\beta\gamma\delta} \widetilde{\cal
    C}_{\alpha\gamma,\beta\delta} ({\bm r}, {\bm r}') \\
    \times
    {\bm J}_{\delta\alpha} ({\bm r}', {\bm r})
    \cdot {\bm J}_{\beta\gamma} ({\bm r}', {\bm r}), \nonumber
\end{eqnarray}
\begin{eqnarray}\label{sigma_b_def}
    \sigma_b = {\hbar^4 \over 2 \pi m \tau} \int d {\bm r} \int d {\bm
    r}'\int d {\bm r}^{\prime \prime}
    \sum_{\alpha\beta\gamma\delta\mu\nu}
    {\cal C}_{\alpha\gamma,\beta\delta} ({\bm r}, {\bm r}') \\
    \times
    \Biggl[
    J^x_{\delta\mu} ({\bm r}', {\bm r}^{\prime \prime})
    J^x_{\nu\gamma} ({\bm r}^{\prime \prime}, {\bm r})
    G^A_{\beta\nu}({\bm r}', {\bm r}^{\prime \prime})
    G^A_{\mu\alpha}({\bm r}^{\prime \prime}, {\bm r})
    \nonumber\\
    +
    J^x_{\beta\nu} ({\bm r}', {\bm r}^{\prime \prime})
    J^x_{\mu\alpha} ({\bm r}^{\prime \prime}, {\bm r})
    G^R_{\delta\mu}({\bm r}', {\bm r}^{\prime \prime})
    G^R_{\nu\gamma}({\bm r}^{\prime \prime}, {\bm r})
    \Biggr]. \nonumber
\end{eqnarray}
Appearance of the modified Cooperon $\widetilde{\cal C} = {\cal C}
- {\hbar^3 \over m \tau} P$ follows from that only three and more
scattering events contribute to the magnetoconductivity.

The current vertex is defined as
$$
    {\bm J} ({\bm r}, {\bm r}') = e \int d {\bm r}_1
    G^R({\bm r}, {\bm r}_1) {\bm v}({\bm
    r}_1) G^A({\bm r}_1, {\bm r}'),
$$
where $\bm v$ is the velocity operator in a magnetic field.
Substituting here the Green functions in the form Eq.~(\ref{G}),
one can show for low magnetic fields Eq.~(\ref{field_range}) that
\begin{equation}\label{J_via_G}
    J^\pm_{\alpha\beta} ({\bm r}, {\bm r}') = {{\rm i} e l \over \hbar} \: e^{\pm {\rm i} \theta}
    \left[ G^R_{\alpha\beta}({\bm r}, {\bm r}') + G^A_{\alpha\beta}({\bm r}, {\bm r}')
    \right],
\end{equation}
where $J^\pm = J^x \pm {\rm i} J^y$, and $\theta$ is an angle
between ${\bm r}$ and ${\bm r}'$.

Omitting the rapidly oscillating products $G^R G^R$ and $G^A G^A$
and expanding the terms $$K({\bm r}, {\bm r}') =  {\rm i}
\cos{\theta} \: P({\bm r}, {\bm r}')$$ in series
Eq.~(\ref{P_expand}), we get from
Eqs.~(\ref{sigma_a_def})-(\ref{J_via_G}) the final expressions for
the conductivity corrections
\begin{equation}\label{sigma_a_fin}
    \sigma_a = - {e^2 \over 2 \pi^2 \hbar} \left( {l \over l_B}
    \right)^2\sum_{N=0}^\infty \Biggl\{ {\rm Tr} \left[ A_N^3 (I - A_N)^{-1} \right]
    - {P_N^3 \over 1 - P_N}\Biggr\},
\end{equation}

\begin{eqnarray}\label{sigma_b_fin}
    \sigma_b = {e^2 \over 4 \pi^2 \hbar} \left( {l \over l_B}
    \right)^2 \sum_{N=0}^\infty
    \Biggl\{ {\rm Tr}
    \left[ K_N \widetilde{K}_N A_N (I - A_N)^{-1} \right] \\
     + {\rm Tr}
     \left[ \widetilde{K}_N K_N A_{N+1} (I - A_{N+1})^{-1} \right] \nonumber \\
   - Q_N^2\left({P_N \over 1 - P_N} + {P_{N+1} \over 1 - P_{N+1}}\right)
   \Biggr\}. \nonumber
\end{eqnarray}
The terms with matrices here are the triplet contributions which
is seen to be of opposite sign in comparison to the singlet ones.
The matrices $K_N$ and $\widetilde{K}_N$ appearing in the
expansion of the function $K({\bm r}, {\bm r}')$ are given by
\begin{equation}\label{K_N}
    K_N = \left(%
\begin{array}{ccc}
  Q_{N-2} - S_{N-2}^{(1)} & R_{N-2}^{(2)} & S_{N-2}^{(3)} \\ \nonumber
  -R_{N-1}^{(0)} & Q_{N-1} - 2S_{N-1}^{(1)} & R_{N-1}^{(2)} \\ \nonumber
  -S_{N-1}^{(1)} & -R_{N}^{(0)} & Q_N - S_N^{(1)} \\
\end{array}%
\right),
\end{equation}
\begin{equation}\label{K_tilde_N}
    \widetilde{K}_N = \left(%
\begin{array}{ccc}
  Q_{N-2} - S_{N-2}^{(1)} & -R_{N-1}^{(0)} & S_{N-1}^{(1)} \\ \nonumber
  -R_{N-2}^{(2)} & Q_{N-1} - 2S_{N-1}^{(1)} & -R_N^{(0)} \\ \nonumber
  S_{N-2}^{(3)} & -R_{N-1}^{(2)} & Q_N - S_N^{(1)} \\
\end{array}%
\right),
\end{equation}
where
\begin{eqnarray}
    &Q_N&= {1 \over \sqrt{N+1}} {l_B \over l} \nonumber\\
    &\times& \int\limits_0^\infty dx
    \exp{\left( -x {l_B \over l} - {x^2 \over 2}\right)}
    x L_N^1(x^2).\nonumber
\end{eqnarray}
Note that the values with negative indices appearing in
Eqs.~(\ref{A_N}),~(\ref{K_N}), and~(\ref{K_tilde_N}) at $N=0,1$
should be replaced by zeros.

Eqs.~(\ref{sigma_a_fin}) and~(\ref{sigma_b_fin}) yield the
weak-antilocalization correction to the conductivity in the whole
range of classically-weak magnetic fields and for arbitrary values
of $\Omega \tau$.

\section{Limiting cases}
In the limit of zero spin splitting, $S_N^{(m)}=R_N^{(m)}=0$, the
matrices $A_N$, $K_N$, and $\widetilde{K}_N$ became diagonal, and
we obtain
\begin{equation}\label{sigma_a_no_spin}
    \sigma_a = - {e^2 \over \pi^2 \hbar} \left( {l \over l_B}
    \right)^2\sum_{N=0}^\infty {P_N^3 \over 1 - P_N},
\end{equation}
\begin{equation}\label{sigma_b_no_spin}
    \sigma_b = {e^2 \over 2 \pi^2 \hbar} \left( {l \over l_B}
    \right)^2\sum_{N=0}^\infty
    Q_N^2 \left({P_N \over 1 - P_N} + {P_{N+1} \over 1 - P_{N+1}}\right).
\end{equation}
Eqs.~(\ref{sigma_a_no_spin}) and~(\ref{sigma_b_no_spin}) coincide
with the results of non-diffusive theory developed for $\Omega =
0$ in Ref.\onlinecite{Zyuzin}.

In the diffusion regime, when $B \ll B_{tr}$, one can calculate
the difference between the conductivity in the presence and in the
absence of magnetic field, $\Delta\sigma(B)$. Making use of
standard approximations valid in the diffusion regime, we obtain
from Eq.~(\ref{sigma_a_fin})
\begin{equation}\label{sigma_diff}
    \Delta\sigma_{diff}(B) =  {e^2 \over 4 \pi^2 \hbar}
    \left[ {\zeta \over \xi} F_T(B) - F_S (B) \right].
\end{equation}
Here the singlet contribution is given by
\begin{equation}\label{Fs}
  F_S(B) = \Psi(1/2 + b_\phi) - \ln{b_\phi},
\end{equation}
where $\Psi$ is the digamma-function. The expression for the
triplet term is as follows\cite{diff_B0}
\begin{eqnarray}\label{Ft}
F_T(B) = - {1 \over a_0} - {2 a_0 + 1 + b_s \over a_1(a_0+b_s) - 2
\zeta b_s} + \sum_{N=1}^\infty \Biggl\{ {\xi+2 \over N}
\nonumber \\
- {(\xi+2) a_N^2 + 2 a_N b_s - \xi - 2(2N+1) \zeta b_s \over
(a_N+b_s)a_{N-1}a_{N+1} -
2 \zeta b_s [(2N+1)a_N - 1]} \Biggr\} \nonumber \\
- {\xi+3 \over 2} \ln{(b_\phi + b_s)} - {\xi+1 \over 2}
\ln{(b_\phi + 2 b_s)}. \hspace{0.5cm}
\end{eqnarray}
Here $a_N =  N + 1/2 + b_\phi + b_s$,
\begin{eqnarray}
b_\phi &=& {\tau \over \tau_\phi}\: {B_{tr} \over B}, \:\:\: b_s =
{2 (\Omega \tau)^2 \over 1 + (2 \Omega \tau)^2} \:
{B_{tr} \over B}, \nonumber \\
\xi &=& \left[1 + 2 (\Omega \tau)^2\right]^{-3}, \:\:\: \zeta =
\left[1 + (2\Omega \tau)^2\right]^{-3}. \nonumber
\end{eqnarray}
Equations~(\ref{sigma_diff})-(\ref{Ft}) generalize the diffusion
theory Ref.\onlinecite{ILP} to the case of arbitrary strong
spin-orbit interaction. In the limit $\Omega \tau \ll 1$, these
expressions pass into the results of Ref.\onlinecite{ILP}.

The conductivity correction in zero magnetic field, $\sigma(0)$,
can be obtained from Eqs.~(\ref{sigma_a_fin})
and~(\ref{sigma_b_fin}) by passing from summation over $N$ to
integration and using the following asymptotic valid for $N \gg 1$
$${1 \over \sqrt{N^m}} x^m L_N^m(x^2) \approx J_m(2x\sqrt{N}).$$
As a result, we get for $\tau_\phi \gg \tau$
\begin{eqnarray}\label{zero_field_a}
&&    \sigma_a(0) = - {e^2 \over 4 \pi^2 \hbar}\\
&\times&    \Biggl\{  {1 \over 2}
    \int\limits_0^\infty dx {\rm Tr} \left[ A_x^3 (I - A_x)^{-1} \right]
    - \ln{\tau_\phi \over \tau}
    \Biggr\}, \nonumber
\end{eqnarray}

\begin{eqnarray}\label{zero_field_b}
&& \sigma_b(0) = {e^2 \over 4 \pi^2 \hbar}
\\
&& \times    \Biggl\{  {1 \over 4} \int\limits_0^\infty dx {\rm
Tr} \left[     (K_x \widetilde{K}_x + \widetilde{K}_x K_x) A_x (I
- A_x)^{-1}     \right]
    - \ln{2}
    \Biggr\}.
    \nonumber
\end{eqnarray}
The matrices here are given by
\begin{equation}\label{Ax}
    A_x = \left(%
\begin{array}{ccc}
  P_x - S_x^{(0)} & R_x^{(1)} & S_x^{(2)} \\ \nonumber
  R_x^{(1)} & P_x - 2S_x^{(0)} & R_x^{(1)} \\ \nonumber
  S_x^{(2)} & R_x^{(1)} & P_x - S_x^{(0)} \\
\end{array}%
\right),
\end{equation}

\begin{equation}\label{Kx}
    K_x = \left(%
\begin{array}{ccc}
  Q_x - S_x^{(1)} & R_x^{(2)} & S_x^{(3)} \\ \nonumber
  -R_x^{(0)} & Q_x - 2S_x^{(1)} & R_x^{(2)} \\ \nonumber
  -S_x^{(1)} & -R_x^{(0)} & Q_x - S_x^{(1)} \\
\end{array}%
\right),
\end{equation}

\begin{equation}\label{K_tilde_x}
    \widetilde{K}_x = \left(%
\begin{array}{ccc}
  Q_x - S_x^{(1)} & -R_x^{(0)} & S_x^{(1)} \\ \nonumber
  -R_x^{(2)} & Q_x - 2S_x^{(1)} & -R_x^{(0)} \\ \nonumber
  S_x^{(3)} & -R_x^{(2)} & Q_x - S_x^{(1)} \\
\end{array}%
\right),
\end{equation}
where
$$
    P_x = {1 \over \sqrt{(1 + \tau/\tau_\phi)^2 + x}}, \:\:\:\:
    Q_x = {1 \over \sqrt{x}} \left( 1 - {1 \over \sqrt{1+x}}
    \right),
$$

$$
    S_x^{(m)} = \int\limits_0^\infty dy \exp{(-y)} J_m(y \sqrt{x}) \sin^2{(\Omega \tau y)},
$$

$$    R_x^{(m)} = {1 \over \sqrt{2}} \int\limits_0^\infty dy
\exp{(-y)} J_m(y \sqrt{x}) \sin{(2\Omega \tau
    y)}.
$$
In Refs.\onlinecite{zero_field} $\sigma(0)$ has been analyzed in
the diffusion approximation $\ln{(\tau_\phi / \tau)} \gg 1$ which
is hardly realized practically.

In a magnetic field $B \gg (\Omega \tau)^2 B_{tr}$, the
conductivity becomes independent of $\Omega$. The reason is that
in so strong field the dephasing length due to magnetic field
$\sim l_B$ is smaller than one due to spin-orbit interaction, $l /
\Omega \tau$. As a result, the particle spins keep safe at
characteristic trajectories. The conductivity for any finite
$\Omega\tau$ has the zero-$\Omega$ asymptotic
Eqs.~(\ref{sigma_a_no_spin}),~(\ref{sigma_b_no_spin}). For
$\Omega\tau < 1$ this dependence is achieved at $B \lesssim
B_{tr}$. In high magnetic field $B \gg B_{tr}, (\Omega \tau)^2
B_{tr}$, the conductivity correction has the high-field
asymptotic\cite{Zyuzin}
$$\sigma_{hf}(B) = - 0.25\sqrt{B_{tr} \over B} \: {e^2 \over \hbar}.$$

\section{Results and Discussion}
\begin{figure}
\includegraphics[width=\linewidth]{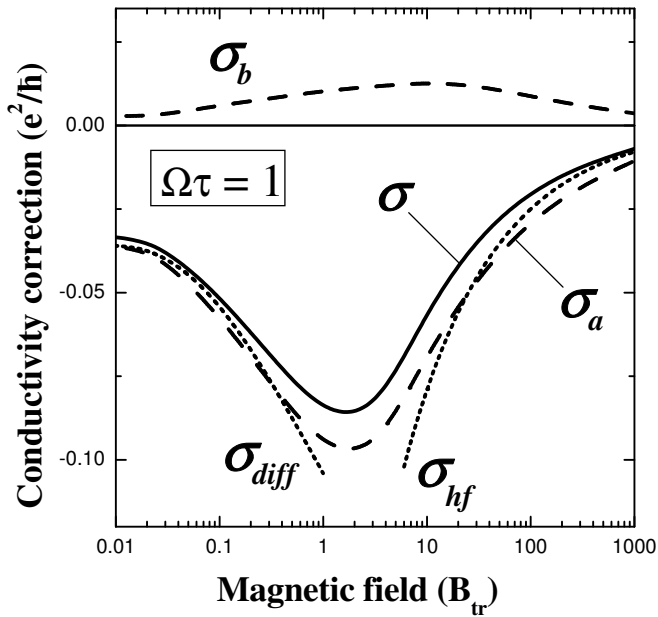}
\caption{Conductivity correction (solid curve) at  $\Omega\tau =
1$, $\tau/\tau_\phi = 0.01$. Dashed curves represent the
backscattering ($\sigma_a$) and nonbackscattering ($\sigma_b$)
contributions, dotted curves show the results of diffusion and
high-field approximations.}
 \label{Wt1}
\end{figure}

In Fig.~\ref{Wt1} weak-antilocalization correction to the
conductivity for $\Omega\tau=1$ is shown by a solid line. In low
fields the conductivity decreases and reaches a minimum at some $B
= B_{min}$. Then the field dependence asymptotically tends to
zero. Dashed curves in Fig.~\ref{Wt1} represent the backscattering
($\sigma_a$) and nonbackscattering ($\sigma_b$) contributions. One
can see that $\sigma_b$ can reach almost 25\% of $|\sigma_a|$,
therefore the nonbackscattering correction should be taken into
account when fitting experimental data. The dotted curves in
Fig.~\ref{Wt1} show results of the diffusion and high-field
approximations.\cite{sigma_diff} One can see that the former is
valid in a narrow region $B < 0.5~B_{tr}$. The high-field
asymptotic holds true only for $B > 100~B_{tr}$. This proves
importance of non-diffusion theory for high-mobility structures.

\begin{figure}
\includegraphics[width=\linewidth]{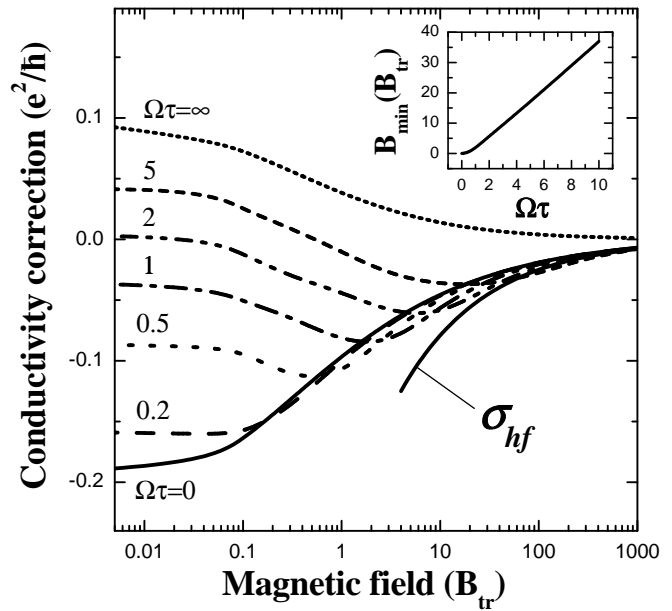}
\caption{Conductivity correction for different strengths of
spin-orbit interaction at $\tau/\tau_\phi = 0.01$. The inset
represents the positions of minima in the magnetoconductivity.}
 \label{cond}
\end{figure}

In Fig.~\ref{cond} the conductivity correction is plotted for
different strengths of spin-orbit interaction. One can see that
for $\Omega\tau \lesssim 1$, in accordance with results of the
previous Section, $\sigma(B)$ coincides with the zero-$\Omega$
dependence for $B > B_{min}$. The asymptotic $\sigma_{hf}(B)$ is
reached at $B \approx 100~B_{tr}$ for all finite values of $\Omega
\tau$. The positions of minima in the curves are shown in the
inset. One can see that $B_{min}$ almost linearly depends on the
spin splitting at $\Omega \tau > 0.8$. Fitting yields the
following approximate law
$$B_{min} \approx (3.9 \: \Omega \tau -2) B_{tr}.$$

In the limit $\Omega\tau \rightarrow \infty$, the triplet state
with $m_s = 0$ does not contribute to the conductivity. The
corresponding dependence is presented in Fig.~\ref{cond}. One can
see a decrease of conductivity in the whole range of magnetic
fields. At $B \gg B_{tr}$, the correction tends to zero as $0.035
\: {e^2 / \hbar} \: \sqrt{B_{tr} / B}$.

\begin{figure}
\includegraphics[width=\linewidth]{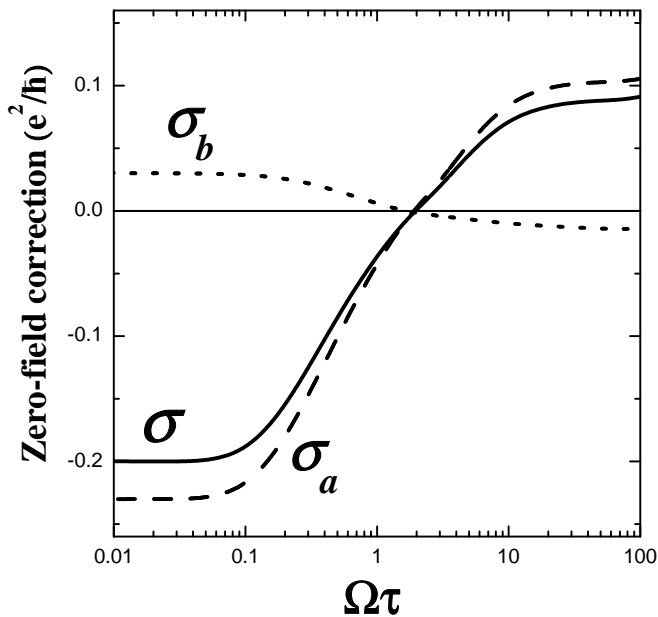}
\caption{Zero-field correction to the conductivity for
$\tau/\tau_\phi = 0.01$ (solid curve). Dashed curves represent the
backscattering ($\sigma_a$) and nonbackscattering ($\sigma_b$)
contributions.}
 \label{zero_field}
\end{figure}

In experiments, the difference $\Delta\sigma(B) =
\sigma(B)-\sigma(0)$ is measured. Since $\sigma(B)$ tends to zero
at $B \to \infty$ for any $\Omega\tau$, one can extract
$\sigma(0)$ from the saturation value of $\Delta\sigma$ at $B \to
\infty$. In Fig.~\ref{zero_field} the zero-field value of the
conductivity correction $\sigma(0)$ is plotted as a function of
$\Omega\tau$.

Fig.~\ref{zero_field} shows how spin-orbit interaction changes the
sign of weak-localization correction to conductivity. At
$\Omega\tau = 0$, all three triplet states and a singlet one yield
contributions of the same absolute value. As a result, the
zero-field correction is given by
\[\sigma_a^0(0) = - {e^2 \over 2 \pi^2 \hbar}
\ln{\tau_\phi \over \tau}, \:\:\:\: \sigma_b^0(0) = {e^2 \over 2
\pi^2 \hbar} \ln{2}.\]
In the opposite limit $\Omega\tau \to \infty$, the triplet
contribution is partially suppressed by the spin-orbit
interaction. Calculation shows that the corrections reach the
following values
\begin{equation}\label{Omega_inf_B_0}
    \sigma_a^\infty(0) = {e^2 \over 4\pi^2 \hbar}
    \left( 0.57 + \ln{\tau_\phi \over \tau}
    \right), \:\:\:
\sigma_b^\infty(0) = -0.43 {e^2 \over 4\pi^2 \hbar}.
\end{equation}
One can see that $\sigma(0)$ changes its sign and reduces its
absolute value when $\Omega \tau$ increases from zero to infinity.

It follows from Fig.~\ref{zero_field} that the magnetoconductivity
$\Delta\sigma(B)$ is an alternating function at small $\Omega
\tau$, while at large values of the spin splitting
$\Delta\sigma(B)$ is negative in the whole range of classically
weak magnetic fields.

\section{Conclusion}

In the present paper the anomalous magnetoresistance is calculated
for 2D systems with only Rashba or only Dresselhaus spin-orbit
interaction. In both cases the spin splitting is isotropic in $\bm
k$-space and characterized by one constant $\Omega$. In the
presence of both types of spin-orbit interaction,
Eqs.(\ref{G}),~(\ref{G_phases}), and~(\ref{C_Eq})-(\ref{C_S}) hold
true with ${\bm \omega} = {\bm \omega}_R + {\bm \omega}_D$.
However $P_T(N,N')$ is not divided into finite blocks as
Eq.(\ref{A_N}), and one should use an infinite matrix for
calculation of the triplet contribution to the conductivity in
this case.

The problem has an analytic solution if the Rashba and Dresselhaus
spin splittings are equal to each other. In this case the
magnetoconductivity is positive in the whole range of magnetic
fields like in systems without spin-orbit interaction. This result
has been previously obtained in the diffusion approximation for $B
\ll B_{tr}$.\cite{PikusPikus} For magnetic field of arbitrary
strength, the dependence $\sigma(B)$ is given by
Eqs.(\ref{sigma_a_no_spin}),~(\ref{sigma_b_no_spin}). This can be
proved by noting that at $\Omega_R = \Omega_D$ the vector ${\bm
\Omega}({\bm k})$ is directed along the same axis
for all $\bm k$. As a result, the energy spectrum consists of two
identical paraboloids shifted relative to each other in the
direction of $\bm \Omega$.\cite{STNS} Both these spin subbands
independently yield equal conductivity corrections coinciding with
those for spinless case. The same result takes place for
symmetrical [110]- and [113]-grown quantum wells.

Application of an in-plane magnetic field ${\bm B}_\parallel$
destroys weak antilocalization. It has been demonstrated
experimentally that the magnetoconductivity minimum disappears in
the presence of ${\bm B}_\parallel$.\cite{tilted,Nitta_B_par} Weak
antilocalization in a tilted magnetic field can be also described
by the present theory. Parallel field influences the anomalous
magnetoresistance due to two microscopic reasons. First, an
in-plane field results in additional dephasing due to orbital
effects.\cite{Malsh1,Meyer} They can be taken into account as
$B_\parallel$-dependent corrections to $\tau_\phi$. Second, an
in-plane field induces finite Zeeman splitting. This results in a
mixing of the singlet and triplet states,\cite{Malsh1} which makes
the matrix $P(N,N')$ in Eq.(\ref{C_coeff}) infinite. However if
the Zeeman splitting is much smaller than $\hbar \Omega$, then
$B_\parallel$ affects only the singlet state. It leads to another
correction to $\tau_\phi$ which should be taken into account only
in ${\cal C}_S$, Eq.(\ref{C_S}). Both the Zeeman and the orbital
corrections to the dephasing rate can be extracted from the fit of
experimental data by Eqs.(\ref{sigma_a_fin}),~(\ref{sigma_b_fin}).
Inclusion of the dephasing corrections into
Eqs.(\ref{zero_field_a}),~(\ref{zero_field_b}) allows one to
describe anomalous magnetoresistance in pure in-plane field as
well.

In conclusion, the theory of weak antilocalization is developed
for high-mobility 2D systems. Anomalous magnetoconductivity is
calculated in the whole range of classically weak fields and for
arbitrary values of spin-orbit splitting.

\acknowledgments{Author thanks S.\,A.~Tarasenko and M.\,M.~Glazov
for discussions. This work was financially supported by the RFBR,
INTAS, and ``Dynasty'' Foundation --- ICFPM.}


\begin{thebibliography}{20}
\bibitem{diff_B0}The zero-field value $F_T(0)$ is a small constant
which should be extracted from $\Delta \sigma_{diff}$ as it has
been done in Ref.~\onlinecite{tilted}.
\bibitem{sigma_diff}$\sigma_{diff}(B)$ is calculated as $\Delta \sigma_{diff}(B) + \sigma(0)$, with use of Eqs.~(\ref{sigma_diff}) and~(\ref{zero_field_a}),~(\ref{zero_field_b}).
\end{thebibliography}
\end{document}